\documentclass[preprint,aps,amssymb,showpacs,superscriptaddress,nofootinbib]{revtex4}
\usepackage{graphicx}
\newcommand{\del}{\partial}

\newcommand{\Case}[2]{{\textstyle \frac{#1}{#2}}}

\begin{document}

%\today
\preprint{IMSc/2008/11/16}

\title{Topological Interpretation of Barbero-Immirzi Parameter }

\author{Ghanashyam Date}
\email{shyam@imsc.res.in}
\affiliation{The Institute of Mathematical Sciences\\
CIT Campus, Chennai-600 113, INDIA.}

\author{Romesh K. Kaul}
\email{kaul@imsc.res.in}
\affiliation{The Institute of Mathematical Sciences\\
CIT Campus, Chennai-600 113, INDIA.}

\author{Sandipan Sengupta}
\email{sandi@imsc.res.in}
\affiliation{The Institute of Mathematical Sciences\\
CIT Campus, Chennai-600 113, INDIA.}

\begin{abstract}
We set up a canonical Hamiltonian formulation for a theory of gravity
based on a Lagrangian density made up of the Hilbert-Palatini term and,
instead of the Holst term, the Nieh-Yan topological density.  The
resulting set of constraints in the time gauge are shown to lead to a
theory in terms of a real $SU(2)$ connection which is exactly the same
as that of Barbero and Immirzi with the coefficient of the Nieh-Yan term
identified as the inverse of Barbero-Immirzi parameter. This provides a
topological interpretation for this parameter. Matter coupling can then
be introduced in the usual manner, {\em without} changing the universal
topological Nieh-Yan term.
\end{abstract}


\maketitle

%%%%%%%%%%%%%%%%%%%%%%%%%%%%%%%%%%%%%%%%%%%%%%%%%%%%%%%%%%%%%%%%%%%%%%%

\section{Introduction}

Hilbert-Palatini Lagrangian for pure gravity is written in terms of the
connection fields $ \omega_{\mu}^{IJ}$ and tetrad $e^{I}_{\mu}$
as independent field variables. Its Holst generalisation is given in
terms of the Lagrangian density \cite{Holst}:
\begin{equation} \label{LagrangianZero}
{\cal L} ~ = ~ \frac{1}{2}e\Sigma^{\mu\nu}_{IJ}R_{\mu\nu}^{~~~ IJ}(\omega) ~ + ~
\frac{\eta}{2}e\Sigma^{\mu\nu}_{IJ}\tilde{R}_{\mu\nu}^{~~~ IJ}(\omega)
\end{equation}
where,
\[
\Sigma_{IJ}^{\mu\nu} ~ := ~
\frac{1}{2}(e_{I}^{\mu}e_{J}^{\nu}-e_{J}^{\mu}e_{I}^{\nu})  ~,~  
R^{~~~ IJ}_{\mu\nu}(\omega) ~ := ~
\partial_{[\mu} \omega_{\nu]}^{~IJ} +
\omega_{[\mu}^{~IK}\omega_{\nu]K}^{~ ~ ~J}  ~,~  
\tilde{R}_{\mu\nu}^{~~~ IJ}(\omega) ~ := ~
\frac{1}{2}\epsilon^{IJKL}R_{\mu\nu KL}(\omega).
\]

The second term is the Holst term with $\eta^{-1}$ as the
Barbero-Immirzi parameter \cite{FernandoBarbero,Immirzi}. For $\eta =
-i$, this Lagrangian density leads to the canonical formulation in terms
of the self-dual Ashtekar connection which is a {\em complex} $SU(2)$
connection \cite{Ashtekar}. For real $\eta$, we have a Hamiltonian
formulation in terms of a {\em real} $SU(2)$ connection, which coincides
with the Barbero formulation for $\eta=1$ \cite{FernandoBarbero,Sa}.

Inclusion of Holst term does not change the classical equation of motion
of the Hilbert-Palatini action; there is no dependence on $\eta$ in the
equations of motion. In fact, when the connection equation
$\omega_{\mu}^{~IJ} = \omega_{\mu}^{~IJ}(e)$ is used, Holst term is
identically zero.

Adding matter in the generalised Lagrangian density
(\ref{LagrangianZero}) needs special care. In particular when spin
$\Case{1}{2}$ fermions are included through minimal coupling, the
classical equations of motion acquire a dependence on
$\eta~$\cite{Freidel}. However it is possible to modify the Holst term
in such a way that the equations of motion remain unchanged.  Such
modification for spin $\Case{1}{2}$ fermionic matter and also those in
the $N=1, 2$ and $4$ supergravities have been obtained
\cite{Mercuri,Kaul}.  {\em When the connection equation of motion is
used,} the modified Holst terms in each of these cases, become total
divergences involving Nieh-Yan invariant density and divergence of axial
current densities involving the fermion fields. The modified Holst term
used in these formulations {\em changes} with the matter content of the
theory.

It has been suggested that the Barbero-Immirzi parameter should have a
topological interpretation in the same manner as the $\theta$ parameter
of QCD \cite{GambiniPullin}. For this to be the case, $\eta$ should be
the coefficient of a term in the Lagrangian density which is a {\em
topological density}. Since such a term would be a total derivative for
{\em all} field configurations, the classical equations of motion would
remain unaltered. Such a term would be universal in the sense that it
would not change when any matter coupling to gravity is introduced. The
Holst term in (\ref{LagrangianZero}) or any of its modifications
mentioned above do not have such a property.

In the four dimensional gravity, there are three possible topological
densities, namely Pontryagin, Euler and Nieh-Yan.  The first two are
quadratic in the curvature tensor. The Nieh-Yan density contains a term
linear in $R_{\mu\nu}^{~~~ IJ}(\omega)$ and an $R-$independent term.
This is shown below to be associated with Barbero-Immirzi parameter.

The Nieh-Yan density is given by \cite{Nieh}:
\begin{equation}
I_{NY} ~ = ~
\epsilon^{\mu\nu\alpha\beta}\left[D_{\mu}(\omega)e^{I}_{\nu}\
D_{\alpha}(\omega)e_{I\beta} - \frac{1}{2}\Sigma^{IJ}_{\mu\nu}\
R_{\alpha\beta IJ}(\omega)\right] ~ ~,~ ~ D_{\mu}(\omega)e_{\nu}^I ~ :=
~ \partial_{\mu}e_{\nu}^I + \omega_{\mu~ J}^{~I}e_{\nu}^J \ .
\end{equation}
This is a topological density, that is, it is a total divergence:
\begin{equation} \label{JDefn}
I_{NY} ~ = ~ \partial_{\mu}J^{\mu}_{\mathrm{NY}}(e, \omega) ~ ~ , ~ ~
J^{\mu}_{\mathrm{NY}}(e, \omega) ~ := ~
\epsilon^{\mu\nu\alpha\beta}e^{I}_{\nu}D_{\alpha}(\omega)e_{I\beta} \ .
\end{equation}
Note that unlike the Pontryagin and Euler densities, the Nieh-Yan
density {\em vanishes identically for a torsion free connection}.

The classical equations of motion from the Lagrangian density containing
the Hilbert-Palatini term as well as the Nieh-Yan density:
\begin{equation} \label{LagrangianOne}
{\cal L} ~ = ~ \frac{1}{2}e\Sigma^{\mu\nu}_{IJ}R_{\mu\nu}^{~~~
IJ}(\omega)+\frac{\eta}{2}I_{NY}
\end{equation}
are the same as those from the Hilbert-Palatini Lagrangian alone.  We
shall demonstrate that the canonical Hamiltonian formulation based on
this new Lagrangian density also leads to a theory of real $SU(2)$
connections, exactly the same as that emerging from the theory with
original Holst term. This in turn, for $\eta = 1$, is the Barbero
formulation. Inclusion of matter now does not need any further
modification and equations of motion continue to be independent of
$\eta$ for all couplings.  This also allows a direct interpretation of
the $\eta$ parameter as a topological parameter in a manner analogous to
the $\theta$-parameter in QCD. 

In a quantum framework, it is also possible to arrive at the canonical
formulation based on the Lagrangian density (\ref{LagrangianOne})
starting from the Hilbert-Palatini canonical formulation by rescaling
the wave functional by exp$\{\Case{i\eta}{2}\int d^3x
J^t_{\mathrm{NY}}(e, \omega)\}$. Mercuri has used this approach to
derive the canonical formulation containing the Barbero-Immirzi
parameter for a theory with spin-1/2 fermions \cite{MercuriNY}. This
demonstrated for the first time, the role of Nieh-Yan density as the
source of the quantization ambiguity reflected by the Barbero-Immirzi
parameter. However, in this analysis the connection equation of motion
has been used to express the $J^t_{\mathrm{NY}}(e, \omega)$ in terms of
the fermions.  It is desirable to carry out this procedure, retaining
the $J^t_{\mathrm{NY}}$ as in equation (\ref{JDefn}) in terms of the
original geometric variables. Such a method then can be applied directly
to a theory of gravity with or without matter. 

In this paper, we work within a classical framework. In section II, we
describe the Hamiltonian formulation based on the Lagrangian density
(\ref{LagrangianOne}) closely following the analysis carried out by Sa
\cite{Sa} for the Hilbert-Palatini gravity with the Holst term. In
section III, we discuss the matter couplings, in particular the case of
Dirac fermions. Coupling of any other matter can be done in an analogous
and straight forward manner. Section IV contains a few concluding
remarks.

\section{Hamiltonian Analysis}

We propose the Lagrangian density for pure gravity to be that given in
equation (\ref{LagrangianOne}), rewritten as: 
\begin{eqnarray}
{\cal L} & = & \frac{1}{2}e\Sigma^{\mu\nu}_{IJ}R_{\mu\nu}^{~~~
IJ}(\omega) + \frac{\eta}{2}\left[
e\Sigma^{\mu\nu}_{IJ}\tilde{R}_{\mu\nu}^{~~~ IJ}(\omega) +
\epsilon^{\mu\nu\alpha\beta} D_{\mu}(\omega)e^{I}_{\nu}\
D_{\alpha}(\omega)e_{I\beta}\right] \nonumber \\
& = & \frac{1}{2}e\Sigma^{\mu\nu}_{IJ}R_{\mu\nu}^{(\eta)IJ}(\omega) +
\frac{\eta}{2}\ \epsilon^{\mu\nu\alpha\beta}
D_{\mu}(\omega)e^{I}_{\nu}D_{\alpha}(\omega)e_{I\beta} ~ ~ ~,
\end{eqnarray}
where  $R_{\mu\nu}^{(\eta)IJ}(\omega) ~ := ~ R_{\mu\nu}^{~~~
IJ}(\omega)+\eta\tilde R_{\mu\nu}^{~~~ IJ}(\omega)$ and we have used the
identities,
\begin{equation}
\Sigma^{\mu\nu}_{IJ}\tilde{R}_{\mu\nu}^{~~~ IJ}(\omega) ~ = ~
\tilde\Sigma^{\mu\nu}_{IJ}R_{\mu\nu}^{~~~ IJ}(\omega)~~~,~~~
e\tilde\Sigma^{\mu\nu}_{IJ} ~ := ~
\frac{e}{2}\epsilon_{IJKL}\Sigma^{\mu\nu KL} ~ = ~
-\frac{1}{2}\epsilon^{\mu\nu\alpha\beta}\Sigma_{\alpha\beta IJ} \ .
\end{equation}
Introducing the notation, $t^{a}_{I} :=
\eta\epsilon^{abc}D_{b}(\omega)e_{Ic}$ and $\epsilon^{abc} :=
\epsilon^{tabc}$, the 3+1 decomposition is expressed as:
\begin{eqnarray}
{\cal L} & = & e\Sigma^{ta}_{IJ}R_{ta}^{(\eta)IJ}(\omega)
+\frac{e}{2}\Sigma^{ab}_{IJ}R_{ab}^{(\eta)IJ}(\omega)
+t^{a}_{I}\left(D_{t}(\omega)e^{I}_{a}-D_{a}(\omega)e^{I}_{t}\right)
\end{eqnarray} 
Defining $\omega_a^{(\eta)IJ} := \omega_a^{IJ} + \eta
\tilde{\omega}_a^{IJ}$ and $\Sigma_{IJ}^{(\eta)ta} := \Sigma_{IJ}^{ta} +
\eta \tilde{\Sigma}_{IJ}^{ta}$ we get,
\begin{eqnarray}
{\cal L} & = & e\Sigma^{ta}_{IJ}\partial_{t}\omega_{a}^{(\eta)IJ} +
\omega_{t}^{IJ}D_{a}(\omega) \left(e\Sigma^{(\eta)ta}_{IJ}\right) +
\frac{e}{2}\Sigma^{ab}_{IJ} R_{ab}^{(\eta)IJ}(\omega) \nonumber \\ & &
\hspace{1.0cm} + t^{a}_{I}\partial_{t}e_{a}^{I} + \omega_t^{~IJ}t^a_I
e_{a J} + e_t^{\, I} D_a(\omega) t^a_I -
\partial_{a}\left(t^{a}_{I}e^{I}_{t} + e
\Sigma^{(\eta)ta}_{IJ}\omega_{t}^{IJ}\right)
\end{eqnarray} 
We parametrize the tetrad fields as: 
\begin{equation}
e^{I}_{t} ~ = ~ \sqrt{eN}M^{I}+N^{a}V_{a}^{I} ~ , ~ e^{I}_{a} ~ = ~
V^{I}_{a} ~ ~ ~;~ ~ ~ M_{I}V_{a}^{I} ~ = ~ 0 ~ , ~ M_{I}M^{I} ~ = ~ -1
\end{equation} 
and then the inverse tetrad fields are: 
\begin{eqnarray}
e^{t}_{I} ~ = ~ -\frac{M_{I}}{\sqrt{eN}} ~ , ~ e^{a}_{I} ~ = ~
V^{a}_{I}+\frac{N^{a}M_{I}}{\sqrt{eN}}  ~ ~; \nonumber \\
M^{I}V_{I}^{a} ~ := ~ 0 ~,~ V_a^I V^b_I ~ := ~ \delta_a^b ~,~ V_a^I
V^a_J ~ := ~ \delta^I_J + M^IM_J 
\end{eqnarray} 
Defining $ q_{ab} ~ := ~V_{a}^{I}V_{bI}$ and $q := \mathrm{det}q_{ab}\
$, leads to $e := det(e^{I}_{\mu}) = Nq$.  We may thus trade the 16
tetrad fields with the 9 fields $V^{a}_{I} \ (M^{I}V_{I}^{a} = 0)$, the
3 fields $M^{I} \ (M_{I}M^{I} = -1)$ and the 4 fields $N$ and $N^{a}$.  

Next using the identity,
\begin{equation}
\Sigma_{IJ}^{ab} ~ = ~ 2 N e \Sigma^{t[a}_{IK}\Sigma^{b]t}_{JL}\eta^{KL}
+ N^{[a}\Sigma^{b]t}_{IJ} \ ,
\end{equation}
and dropping the total space derivative terms, 
\begin{equation}
{\cal L} =e \Sigma^{ta}_{IJ} \partial_{t}\omega_{a}^{(\eta)IJ} +
t^{a}_{I}\partial_{t}e_{a}^{I} - N H - N^{a} H_{a} - \Case{1}{2}
\omega_{t}^{IJ} G_{IJ} 
\end{equation} 
where $2e\Sigma^{ta}_{IJ} = -\sqrt{q}M_{[I}V_{J]}^{a}\ $, $t^{a}_{I} :=
\eta\epsilon^{abc}D_{b}(\omega)V_{Ic}$ and 
\begin{eqnarray}
H & = &
2e^2\Sigma^{ta}_{IK}\Sigma^{tb}_{JL}\eta^{KL}R_{ab}^{(\eta)IJ}(\omega) -
\sqrt{q}M^{I}D_{a}(\omega) t^{a}_{I}, \label{HZero}\\
H_a & = & e\Sigma^{tb}_{IJ}{R_{ab}^{(\eta)IJ}}(\omega) -
V_{a}^{I}D_{b}(\omega) t^{b}_{I}, \label{HaZero} \\
G_{IJ} & = & -2 D_a(\omega) \left( e \Sigma^{(\eta)ta}_{IJ}\right) -
t^a_{[I}V_{J]}a \ . \label{GIJZero} 
\end{eqnarray}

Introduce the fields, 
\begin{equation}
E^{a}_{i} := 2e\Sigma^{ta}_{0i} ~,~ \chi_{i} := -M_{i}/M^{0} ~,~
A^{i}_{a} ~ := ~ \omega_{a}^{(\eta)0i}-\chi_{j}\omega_{a}^{(\eta)ij} ~,~
\zeta^{i} ~ := ~ -E^{a}_{j}\omega_{a}^{(\eta)ij}.
\end{equation}
In terms of these, we have $2e\Sigma^{ta}_{ij}=-E^{a}_{[i}\chi_{j]}$ and
$e\Sigma^{ta}_{IJ}\partial_{t}\omega_{a}^{(\eta)IJ}=E^{a}_{i}\partial_{t}A^{i}_{a}+\zeta^{i}\partial_{t}\chi^{i}$,
and the Lagrangian density is:
\begin{equation}
{\cal L} ~=~ E^{a}_{i}\partial_{t}A^{i}_{a} +
\zeta^{i}\partial_{t}\chi^{i} + t^{a}_{I}\partial_{t}V_{a}^{I} - NH -
N^{a} H_{a} - \Case{1}{2}\omega_{t}^{IJ}G_{IJ},  
\end{equation}
where, now we need to re-express $H, H_a$ and $G_{IJ}$
($G^i_{\mathrm{boost}} := G_{0i}\, ,\  G^i_{\mathrm{rot}} := \Case{1}{2}
\epsilon^{ijk} G_{jk}$) in terms of these new fields: 
\begin{eqnarray}
G^{i}_{ \mathrm {boost}} & = &  -\partial_{a}\left( E^{a}_{i} - \eta
\epsilon^{ijk} E^a_j \chi_k\right) + E^a_{[i}\chi_{k]} A^k_a + (\zeta^i
- \chi\cdot\zeta \chi^i) - t^a_{[0}V_{i]a} \ , \\ \label{GboostOne}
G^{i}_{\mathrm{rot}} & = & \partial_a\left(\epsilon^{ijk}E^{a}_{j}
\chi_{k} + \eta E^a_i\right) + \epsilon^{ijk}\left( A_{a}^{j}E^{a}_{k} -
\zeta_{j}\chi_k -t^a_j V^k_a \right) \\ \label{GrotOne}
H_{a} & = & E^{b}_{i}\left[R_{ab}^{(\eta) 0i}(\omega) -
\chi_{j}R^{(\eta) ij}_{ab}(\omega)\right]  - V_a^I D_b(\omega) t^b_I \\
& = & E^b_i \partial_{[a}A_{b]}^i + \zeta^i \partial_a \chi^i - V_a^I
\partial_b t^b_I + t^b_I \partial_{[a}V_{b]}^I  \nonumber \\
& & - \frac{1}{1 + \eta^2}\left[ E^b_{[i} \chi_{l]} A^l_b + (\zeta_i -
\chi\cdot\zeta\chi_i) - t^b_{[0}V_{i]b} -\eta\epsilon^{ijk}(A^j_b E^b_k
- \zeta_j\chi_k - t^b_jV^k_b)\right] A^i_a \nonumber \\
& & - \frac{1}{1 + \eta^2}\left[ \frac{1}{2}\epsilon^{ijk}(\eta
G^k_{\mathrm{boost}} + G^k_{\mathrm{rot}}) - \chi^i(G^j_{\mathrm{boost}}
- \eta G^j_{\mathrm{rot}})\right] \omega_a^{(\eta)ij} \\ \label{HaOne}
H & = & - E^{a}_{k} \chi_{k} H_{a} - \frac{1}{2}(1-\chi \cdot\chi)
E^{a}_{i}E^{b}_{j} R_{ab}^{(\eta)ij}(\omega) - \left(E^{a}_{k}\chi_{k}
V_{a}^{I} + \sqrt{q} M^{I}\right) D_{b}(\omega) t^{b}_{I} \nonumber \\
& = & - E^a_k \chi_k H_a + (1 - \chi\cdot\chi)\left[ E^a_i\partial_a
\zeta_i + \frac{1}{2} \zeta_i E^a_i E^b_j \partial_a E^j_b \right]
\nonumber \\
& & + \frac{1 - \chi\cdot\chi}{2 (1 + \eta^2)} \zeta_i \left[ -
G^i_{\mathrm{boost}} + \eta G^i_{\mathrm{rot}}\right] - \left( E^a_k
\chi_k V_a^I + \sqrt{q} M^I \right) \partial_b t^b_I \nonumber \\
& & - \frac{1 - \chi\cdot\chi}{1 + \eta^2} \left[ \frac{1}{2} E^a_{[i}
E^b_{j]} A^i_a A^j_b + E^a_i A^i_a \chi\cdot\zeta +
\eta\epsilon^{ijk}\zeta_i A^j_a E^a_k \right. \nonumber \\
& & \left. \hspace{3.0cm} + \frac{3}{4} (\chi\cdot\zeta)^2 - \frac{3}{4}
(\zeta\cdot\zeta) + \frac{1}{2}\zeta_i t^a_{[0}V_{i]a} - \frac{\eta}{2}
\zeta_i \epsilon^{ijk}t^a_j V^k_a \right] \nonumber \\
& & + \frac{1 - \chi\cdot\chi}{1 + \eta^2} \left[\frac{1}{\sqrt{E}}A^i_a
t^a_i + \frac{1}{2} V^i_a \left( \zeta\cdot\chi t^a_i -
\chi_i\zeta_jt^a_j + \eta \epsilon^{ijk} \zeta_j t^a_k \right) \right]
\nonumber \\
& & + \frac{1 - \chi\cdot\chi}{1 + \eta^2} \left[-
\frac{1}{\sqrt{E}}\chi_i t^b_j + \frac{\eta}{2
\sqrt{E}}\epsilon^{ijk}t^b_k + (1 + \eta^2) E^a_i \partial_a E^b_j +
E^a_i\chi_j E^b_m A^m_a \right. \nonumber \\
& & \left. \hspace{3.0cm} - \eta\epsilon^{imn}E^a_mE^b_nA^j_a -
\frac{\eta}{4}\left( \epsilon^{ijm}E^b_n +
\epsilon^{ijn}E^b_m\right)\chi_m\zeta_n\right] u_b^{ij} \nonumber \\
& & + \frac{1 - \chi\cdot\chi}{2(1 + \eta^2)} \left(\chi_m\chi_n -
\delta_{mn}\right) E^a_j E^b_i u^{im}_a u_b^{jn} \label{HOne}
\end{eqnarray}

In the above, $E^{i}_{a} := \sqrt{E}V_a^i$ is the inverse of $E^{a}_{i}$
i.e.  $E^{i}_{a}E^{b}_{i} = \delta^{b}_{a}\, ,\ E^{i}_{a}E^{a}_{j} =
\delta^{i}_{j}$ and $E^{-1} = q (M^0)^2$ equals $\mathrm{det}
E^{a}_{i}$.  Furthermore we have also set $u_a^{ij} :=
\omega_a^{(\eta)ij} - \Case{1}{2} E_a^{[i} \zeta^{j]}$. Notice that
$E^b_i u^{ij}_b = 0$.  The six independent fields in $u^{ij}_a$ may be
parametrized in terms a symmetric matrix $M^{ij}$ as, $u_a^{ij} :=
\Case{1}{2} \epsilon^{ijk}E^l_a M^{kl}$ \cite{Sa}.

We have replaced the original 16 tetrad fields with 16 new fields:
$E^{a}_{i}, \chi_{i}$, $N$ and $N^{a}$. In place of the original 24
connection fields $\omega_{\mu}^{IJ}$ we use the new set of 24 fields
$A^{i}_{a},\zeta_{i} , M^{kl},\omega_{t}^{ij} $ and $\omega_{t}^{0i}$.
The fields $V^{I}_{a}$ and $t^{a}_{I}$ are not independent; these are
given in terms of the fundamental fields as:
$V^{I}_{a}=\upsilon^{I}_{a}$ and $t^{a}_{I}=\tau^{a}_{I}$ where
\begin{eqnarray} \label{UpsilonDefn}
\upsilon^{0}_{a} & := & -\frac{1}{\sqrt{E}}E_{a}^{i}\chi_{i} ~ ~ , ~ ~ 
\upsilon^{i}_{a} ~ := ~ \frac{1}{\sqrt{E}}E_{a}^{i}  \\
\tau^{a}_{0} & := & \eta \epsilon^{abc} D_{b}(\omega)V_{0c} = \eta
\sqrt{E}E^{a}_{m}\left[G^{m}_{\mathrm{rot}}-\frac{\chi_{l}}{2}\left(\frac{2f_{ml}+N_{ml}}{1
+ \eta^2} + \epsilon_{mln}G^{n}_{\mathrm{boost}}\right)\right]
\label{Tau0Defn} \\
\tau^{a}_{k} & := & \eta \epsilon^{abc} D_{b}(\omega)V_{ck} =
-\frac{\eta}{2} \sqrt{E}E^{a}_{m}\left[\frac{2f_{mk} + N_{mk}}{1 +
\eta^2} +\epsilon_{kmn}G^{n}_{\mathrm{boost}}\right] \label{TauADefn}
\end{eqnarray}
where,
\begin{eqnarray}
2f_{kl} & := & \epsilon_{ijk}E^{a}_{i}\left[( 1 + \eta^2 )
E_{b}^{l}\partial_{a}E^{b}_{j} + \chi_{j}A^{l}_{a}\right] + \eta
\left(E^{a}_{l}A_{a}^{k} - \delta ^{kl}E^{a}_{m} A^{m}_{a} -
\chi_{l}\zeta _{k}\right) + (l \leftrightarrow k) \label{FDefn}\\
N_{kl} & := & \epsilon^{ijk}(\chi_{m}\chi_{j} -
\delta_{mj})E^{a}_{i}u_{a}^{lm} + (l \leftrightarrow k) \label{NDefn} \\
& = & (\chi\cdot\chi - 1)( M_{kl}-M_{mm}\delta_{kl} ) + \chi_{m}\chi_{n}
M_{mn}\delta_{kl} + \chi_{l}\chi_{k} M_{mm} - \chi_{m}(\chi_{k} M_{ml} +
\chi_{l} M_{mk}) \nonumber
\end{eqnarray}

We can upgrade $V_{a}^{I}$ and  $t_{I}^{a}$ as independent fields
through terms containing the Lagrange multiplier fields $\xi^{a}_{I}$
and $\phi^{I}_{a}$ in the Lagrangian density:
\begin{eqnarray}
{\cal L} & = &
E^{a}_{i}\partial_{t}A^{i}_{a}+\zeta^{i}\partial_{t}\chi^{i}+t^{a}_{I}\partial_{t}V_{a}^{I}
- {\cal H} \nonumber \\
{\cal H} & := & N H + N^{a} H_{a} + \frac{1}{2} \omega_{t}^{IJ}G_{IJ} +
\xi^{a}_{I}(V_{a}^{I}-\upsilon_{a}^{I}) +
\phi^{I}_{a}(t^{a}_{I}-\tau^{a}_{I}) \label{GravLagrangian}
\end{eqnarray}
where $\upsilon_{a}^{I}$ and $\tau_{I}^{a}$ are defined in equations
(\ref{UpsilonDefn} - \ref{TauADefn}).  We have 24 pairs of canonically
conjugate independent field variables $(E^{a}_{i}, A^{j}_{b}),
(\zeta^{i}, \chi^{j}), ({t}^{a}_{I}, V_{a}^{I})$. The remaining fields,
namely, $N, N^{a}, \omega_{t}^{IJ}, \xi^{a}_{I}, \phi^{I}_{a}$ and
$M^{kl}$ have no conjugate momenta since in the Lagrangian their
velocities do not appear.  Preservation of these constraints (vanishing
of the variation of the Hamiltonian with respect to the fields) leads to
the secondary constraints. From the variations with respect to fields
$\omega_{t}^{0i}, \omega_{t}^{ij}, N^a, N, \xi^{a}_{I}$ and
$\phi^{I}_{a}$, we get the constraints:
\begin{eqnarray}
G^{i}_{\mathrm{boost}} ~ \approx ~ 0 ~ ~ , ~ ~ G^{i}_{\mathrm{rot}} ~
\approx ~ 0 & ~ ~ ; ~ ~ & H_{a} ~ \approx ~ 0 ~ ~ , ~ ~ H ~ \approx ~ 0
~;~ \label{GaugeDiffeoHam}\\ 
V_{a}^{I} - \upsilon_{a}^{I} ~ \approx ~ 0 & ~ ~ , ~ ~ &
t^{a}_{I}-\tau^{a}_{I} ~ \approx ~ 0. \label{Additional}
\end{eqnarray}
From the variation with respect to $M^{kl}$ or equivalently
$u_{a}^{ij}$, we get:
\[
\frac{\delta{\cal H}}{\delta M^{kl}}\delta M^{kl} ~ \approx ~
\frac{\delta H}{\delta M^{kl}}\delta M^{kl} ~ = ~ \frac{(1 -
\chi\cdot\chi)}{2( 1 + \eta^2)}[(\eta t^{a}_{k} -
\epsilon^{ijk}\chi_{i}t^{a}_{j})V_{a}^{l} + f_{kl} + \frac{1}{2}
N_{kl}]\delta M^{kl} \approx 0. \nonumber
\]
This leads to
\begin{equation} \label{Extra}
(\eta t^{a}_{k} - \epsilon^{ijk}\chi_{i}t^{a}_{j})V_{a}^{l} + f_{kl} +
\frac{1}{2} N_{kl} + (k \leftrightarrow l) \approx 0 \nonumber
\end{equation}
Using constraints (\ref{Additional}), and the expressions
(\ref{UpsilonDefn} - \ref{TauADefn}) for $\tau^{a}_{I}, \
\upsilon_{a}^{I}$, equation (\ref{Extra}) implies :
\[
\left(\eta \epsilon^{ijk}\chi_{i} + \delta_{kj}\right) \left( 2f_{jl} +
N_{jl}\right) + \eta (1 + \eta^2) \left( \delta^{kl} \chi_m
G^{m}_{\mathrm{boost}} - \chi_{l} G^{k}_{\mathrm{boost}} \right) + (k
\leftrightarrow l) \approx 0
\]
Using (\ref{GaugeDiffeoHam}), this in turn implies the constraint:
\begin{equation}\label{GravMKLConstraint}
2f_{kl} + N_{kl} \approx 0
\end{equation}
where, $f_{kl}$ and $N_{kl}$ are given in (\ref{FDefn}, \ref{NDefn}).
This constraint can be solved for $M_{kl}$. Furthermore it implies, from
the definitions (\ref{Tau0Defn}, \ref{TauADefn}), that $\tau^{a}_{I}
\approx 0$ and hence,
\begin{equation} \label{GravTorsionConstraint}
t^{a}_{I} \approx 0 \ .
\end{equation}

Implementing this constraint then reduces  the Hamiltonian density to 
\begin{equation}
{\cal H} ~ = ~ N H + N^{a} H_{a} + \frac{1}{2} \omega_{t}^{IJ} G_{IJ}
\end{equation}
where now,
\begin{eqnarray}
G^{i}_{\mathrm{boost}} & = & - \partial_{a}\left(E^{a}_{i} - \eta
\epsilon^{ijk}E^{a}_{j} \chi_{k}\right) + E^{a}_{[i}\chi_{k]} A_{a}^{k}
+ (\zeta ^{i} - \chi\cdot\zeta \zeta^{i}) ~ \approx ~ 0 \ , \\
G^{i}_{\mathrm{rot}} & = & \partial_{a}\left(\epsilon^{ijk}E^{a}_{j}
\chi_{k} + \eta E^{a}_{i}\right) + \epsilon^{ijk} (A^{j}_{a}E^a_k -
\zeta_{j} \chi_{k})  ~ \approx ~ 0 \ , \\
H_{a} & = & E^{b}_{i}\partial_{[a}A_{b]}^{i} +
\zeta_{i}\partial_{a}\chi_{i} \\
& & - \frac{1}{1+\eta ^2}\left[E^{b}_{[k}\chi_{l]} A^{l}_{b} + \zeta_{i}
- \chi\cdot\zeta \chi^{i} - \eta \epsilon^{ijk} (A^{j}_{a}E^a_k
-\zeta_{j} \chi_{k}) \right] A^i_a \nonumber \\
& & -\frac{1}{1+\eta ^2}\left[ \frac{1}{2} \epsilon^{ijk}(\eta
G^{k}_{\mathrm{boost}} + G^{k}_{\mathrm{rot}}) -
\chi^{i}(G^{j}_{\mathrm{boost}}-\eta G^{j}_{\mathrm{rot}})
\right]\omega_{a}^{(\eta)ij} ~ \approx~ 0 \ , \nonumber \\
H & = & - E^{a}_{k}\chi_{k}H_{a} + (1 -
\chi\cdot\chi)\left[E^{a}_{i}\partial_{a}\zeta_{i} + \frac{1}{2}
\zeta_{i}E^{a}_{i}E^{b}_{j}\partial_{a}E_{b}^{j}\right] \\ 
& & + \frac{(1 - \chi\cdot\chi)}{2(1 + \eta^2)}\zeta_{i}[ -
G^{j}_{\mathrm{boost}} + \eta G^{j}_{\mathrm{rot}}] \nonumber \\
& & -\frac{(1 - \chi\cdot\chi)}{1 + \eta ^2}\left[ \frac{1}{2}
E^{a}_{[i}E^{b}_{j]} A^{i}_{a}A^{j}_{b}+E^{a}_{i}A^{i}_{a}\chi\cdot\zeta
+ \eta \epsilon_{ijk}\zeta_{i}A_a^{j}E^{a}_{k} +
\frac{3}{4}(\chi\cdot\zeta)^2 - \frac{3}{4}(\zeta\cdot\zeta)\right]
\nonumber \\
& & + \frac{(1 - \chi\cdot\chi)}{2(1 + \eta^2)}\left[f_{kl}M^{kl} +
\frac{1}{4}(\chi\cdot\chi - 1)(M^{kl}M^{kl} - M^{kk}M^{ll}) \right.
\nonumber \\
& & \left. \hspace{3.0cm} + \frac{1}{2} \chi_{k}\chi_{l}(M^{pp}M^{kl} -
M^{kp}M^{lp})\right] ~ \approx~ 0 \ . \nonumber
\end{eqnarray}
In the last equation we have $M^{kl}$ given by the constraint $2f_{kl} +
N_{kl} = 0$, which can be solved as: 
\begin{equation}
(1 - \chi\cdot\chi)M_{kl} = 2 f_{kl} + (\chi_{m}\chi_{n}f_{mn} -
f_{mm})\delta_{lk} + (\chi_{m}\chi_{n}f_{mn} + f_{mm})\chi_{k}\chi_{l} -
2 \chi_{m}(\chi_{l}f_{mk} + \chi_{k}f_{ml})
\end{equation}

This is the same set of equations as those obtained by Sa \cite{Sa} in
his analysis of the action containing Holst term.

We may fix the boost gauge transformations (time gauge) by imposing
$\chi^i \approx 0$ which together with the $G^{i}_{\mathrm{boost}}
\approx 0$ forms a second class pair.  Solving the boost constraint with
$\chi^i = 0$ yields,
\begin{equation}
\zeta_{i} ~ =~ \partial_{a}E^{a}_{i}
\end{equation}
In this gauge we then recover a canonical Hamiltonian formulation in
terms of real $SU(2)$ gauge fields $A_a^i$ which reduces to the Barbero
formulation for $\eta = 1$ \cite{Sa}.

To summarize, like the Holst term, the Nieh-Yan term leads to an $SU(2)$
gauge theoretic formulation. But, it is only the coefficient of the
Nieh-Yan term that has a topological character.
\section{Matter coupling}

As stated earlier, the matter can now be coupled to gravity in a
straight forward manner. As an example, we consider a spin-$\Case{1}{2}$
Dirac fermion with its usual {\em minimal} coupling to gravity. The
Lagrangian density is\footnote{Our Dirac matrices satisfy the Clifford
algebra: $\gamma^I\gamma^J + \gamma^J\gamma^I = 2 \eta^{IJ} ~,~
\eta^{IJ} := \mathrm{diag}( -1, 1, 1, 1)$. The chiral matrix $\gamma_5
:= i \gamma^0\gamma^1\gamma^2\gamma^3 $ and $ \sigma^{IJ} ~ ~ := ~ ~
\frac{1}{4}~[\gamma^{I},\gamma^{J}] .$},
\begin{equation}
{\cal L} ~ = ~ \frac{1}{2}e~\Sigma^{\mu\nu}_{IJ}~R_{\mu\nu}^{~~~
IJ}(\omega) ~ + ~ \frac{\eta}{2}~I_{NY} ~ + ~ \frac{ie}{2}
~[\bar{\lambda}\gamma^{\mu}D_{\mu}(\omega) \lambda ~ - ~
\overline{D_{\mu}(\omega) \lambda}\gamma^{\mu}\lambda]
\end{equation}
where, 
\begin{eqnarray}
D_{\mu}(\omega)\lambda  ~ := ~ \del_{\mu}\lambda ~ + ~
\frac{1}{2}~\omega_{\mu IJ}~\sigma^{IJ}~\lambda ~ ~,~ ~
\overline{D_{\mu}(\omega)\lambda}  ~ := ~  \del_{\mu}\bar{\lambda} ~ -
~\frac{1}{2}~\bar{\lambda}~\omega_{\mu IJ}~\sigma^{IJ} \nonumber 
\end{eqnarray}

Notice that, unlike earlier attempts of setting up a theory of fermions
and gravity with Barbero-Immirzi parameter \cite{Mercuri,Kaul} where the
Holst term was modified to include an additional non-minimal term for
the fermions, the Lagrangian density here containing the Nieh-Yan
density does not require any further modification, just the usual
minimal fermion terms suffice. This is so because the Nieh-Yan term is
topological. 

We expand the fermion terms as 
\begin{eqnarray}
{\cal L}(F) & := &  \frac{ie}{2}
~[\bar{\lambda}\gamma^{\mu}D_{\mu}(\omega) \lambda ~ - ~
\overline{D_{\mu}(\omega) \lambda}\gamma^{\mu}\lambda] \nonumber \\
& = &  \left[\del_{t}\bar{\lambda} \Pi~-~\bar{\Pi} \del_{t}\lambda
\right] ~-~ N H(F) ~ - ~ N^{a}H_{a}(F) ~ - ~ \frac{1}{2}~
\omega_{t}^{IJ}~G_{IJ}(F) 
\end{eqnarray} 
where $~\bar{\Pi}~$ and $~\Pi~$ are canonically conjugate momenta fields
associated with $~\lambda~$ and $~\bar{\lambda}~$ respectively.
Explicitly \footnote{The fermions are Grassmann valued and the
functional differentiation is done on the left factor which accounts for
the signs in the definitions of the conjugate momenta in
(\ref{MomentaDefns}).},
\begin{eqnarray}
\bar{\Pi} & = & - \frac{i e }{2} \bar{\lambda}\gamma^t ~ = ~ \frac{i
\sqrt{q}}{2}M_{I}\bar{\lambda}\gamma^{I} ~ ~, ~ ~ \Pi ~ = ~ - \frac{i
e}{2} \gamma^t \lambda ~ = ~ \frac{i
\sqrt{q}}{2}M_{I}\gamma^{I}{\lambda} \label{MomentaDefns}\\
G^{IJ}(F) & = & \bar{\Pi}\sigma^{IJ}\lambda ~ + ~
\bar{\lambda}\sigma^{IJ}\Pi \\
H_{a}(F) & = & \overline{D_{a}(\omega)\lambda}\, \Pi ~ - ~
\bar{\Pi}D_{a}(\omega)\lambda\\
H(F) & = & (-
2~e~\Sigma^{ta}_{IJ})~\left[\overline{D_{a}(\omega)\lambda}~\sigma^{IJ}~\Pi
~ + ~ \bar{\Pi}~\sigma^{IJ}~D_{a}(\omega)\lambda\right]
\end{eqnarray}

Incorporating these fermionic terms in the pure gravity Lagrangian
density given in equation (\ref{GravLagrangian}), we write the full
Lagrangian density as,
\begin{eqnarray}
{\cal L} ~& = &~E^{a}_{i}~\del_{t}A^{i}_{a} + \zeta^{i}~\del_{t}\chi_{i}
+ t^{a}_{I}\del_{t}V_{a}^{I} + \del_{t}\bar{\lambda}~\Pi -
\bar{\Pi}~\del_{t}\lambda - NH' - N^{a}H'_{a} - \frac{1}{2}
\omega_{t}^{IJ}~G'_{IJ}~\nonumber\\ 
& & -~\xi^{a}_{I}(V_{a}^{I}-\upsilon_{a}^{I}) -
\phi^{I}_{a}(t^{a}_{I}-\tau^{a}_{I}) 
\end{eqnarray}
where now 
\begin{equation}
G'_{IJ} ~=~ G^{IJ} ~+~ G^{IJ}(F) ~ ~ , ~ ~ H'_{a} ~ =~ H_{a} ~+~
H_{a}(F) ~ ~ , ~ ~  H' ~=~ H ~+~ H(F) \ ,
\end{equation}
with $~G^{IJ}$,$~H_{a}~$ and $~H~$ as the contributions from the pure
gravity sector as given by the equations (\ref{HZero} -- \ref{GIJZero})
or equivalently by the equations (\ref{GboostOne} -- \ref{HOne}).

The various quantities above can then be rewritten in terms of the basic
fields as: 
\begin{eqnarray} 
G'^i_{\mathrm{boost}} & = & -\del_{a}(E^{a}_{i} ~-~ \eta
\epsilon^{ijk}E^{a}_{j} \chi_{k}) ~+~ E^{a}_{[i}\chi_{k]}A_{a}^{k} ~+~
(\zeta ^{i} ~-~ \chi\cdot\zeta~\chi^{i}) - t'^{a}_{[0}V_{i]a}
\nonumber\\
& & +~ \left[ \bar{\Pi} ( 1 + i \eta\gamma_5 ) \sigma_{0i} \lambda ~ + ~
\bar{\lambda}( 1 + i \eta \gamma_5 ) \sigma_{0i} \Pi \right] ;
\label{GBoostF}\\ 
G'^{i}_{\mathrm{rot}} & = & \del_{a}(\epsilon^{ijk}E^{a}_{j} \chi_{k}
~+~ \eta E^{a}_{i}) ~+~ \epsilon^{ijk} (A^{j}_{a}E^a_k ~-~ \zeta_{j}
\chi_{k} ~-~ t'^{a}_{j}V_{a}^{k}) \nonumber\\ 
& & +~ \left[ \bar{\Pi} ( i \gamma_5  - \eta ) \sigma_{0i} \lambda ~ + ~
\bar{\lambda}( i \gamma_5  - \eta) \sigma_{0i} \Pi \right] ;
\label{GRotF}\\ 
H'_{a} & = & E^{b}_{i}\del_{[a}A_{b]}^{i} + \zeta_{i}\del_{a}\chi_{i}
~-~ \del_{b}( t'^{b}_{I}V^{I}_{a}) ~+~
t'^{b}_{I}\del_{a}V^{I}_{b}\nonumber \\
& & + \left[\del_{a}\bar{\lambda}(1 ~+~ i\eta\gamma_{5})\Pi ~-~
\bar{\Pi}(1 ~+~ i\eta\gamma_{5})\del_{a}\lambda \right] ~ -
\left[\bar{\lambda}\sigma_{0i}\Pi +
\bar{\Pi}\sigma_{0i}\lambda\right]A^i_a \nonumber \\
& & - \frac{1}{1 + \eta^2}\left[E^{b}_{[i}\chi_{l]}A^{l}_{b} + \zeta_{i}
- \chi\cdot\zeta\chi^{i} - t'^{b}_{[0}V_{i]b} - \eta \epsilon^{ijk}~
(A^{j}_{a}E^{b}_{k} - \chi_j \zeta_{k} -
t'^{b}_{j}V^{k}_{b})\right]A^{i}_{a} \nonumber \\ 
& & -\frac{1}{1 + \eta^2}\left[\frac{1}{2} \epsilon^{ijk}\left(\eta
G'^{k}_{\mathrm{boost}} + G'^{k}_{\mathrm{rot}}\right) -
\chi^{i}(G'^{j}_{\mathrm{boost}} - \eta G'^{j}_{\mathrm{rot}})
\right]\omega_{a}^{(\eta) ij} \label{DiffeoF} \\ 
H' & = & -E^{a}_{k}\chi_{k}H'_{a} - (E^{a}_{k}\chi_{k}V^{I}_{a} +
\sqrt{q}M^{I})~\del_{b} t'^{b}_{I} + (1 -
\chi\cdot\chi)\left[E^{a}_{i}\del_{a}\zeta_{i} + \frac{1}{2}
\zeta_{i}E^{a}_{i}E^{b}_{j}\del_{a}E^{j}_{b}\right]\nonumber \\ 
& & - \frac{(1 - \chi\cdot\chi)}{1 +
\eta^2}\left[\frac{1}{2}E^{a}_{[i}E^{b}_{j]}A^{i}_{a}A^{j}_{b} +
E^{a}_{i}A^{i}_{a}\chi\cdot\zeta + \eta
\epsilon_{ijk}\zeta_{i}A_{a}^{j}E^{a}_{k} +
\frac{3}{4}(\chi\cdot\zeta)^2 - \frac{3}{4}(\zeta\cdot\zeta)\right]
\nonumber \\ 
& & + \frac{(1 - \chi\cdot\chi)}{2(1 + \eta^2)}\left[\zeta_{i} - 2
V^{i}_{b} (t^{b}_{0} - t'^{b}_{0})\right]\left[- G'^{i}_{\mathrm{boost}}
+ \eta G'^{i}_{\mathrm{rot}} - t'^{a}_{[0}V_{i]a} + \eta\epsilon_{ijk}
t'^{a}_{j}V^{k}_{a}\right] \nonumber \\
& & + \frac{(1 - \chi\cdot\chi)}{\sqrt{E}(1 + \eta^2)}\left[
t'^{b}_{m}A^{m}_{b} + \frac{1}{2}E^{i}_{b} t'^{b}_{[i}\chi_{j]}\zeta_{j}
+ \frac{\eta}{2}\epsilon_{ijk}
t'^{b}_{i}E^{j}_{b}\zeta_{k}\right]\nonumber \\ 
& & -~ 2e~\Sigma^{ta}_{IJ}\left[\del_{a}\bar{\lambda}(1 +
i\eta\gamma_{5})\sigma^{IJ}\Pi + \bar{\Pi}(1 +
i\eta\gamma_{5})\sigma^{IJ}\del_{a}\lambda\right]\nonumber \\ 
& & +~ E^{a}_{k}\chi_{k}\left[\del_{a}\bar{\lambda}(1 +
i\eta\gamma_{5})\Pi - \bar{\Pi}(1 +
i\eta\gamma_{5})\del_{a}\lambda\right]\nonumber \\ 
& & -~ 2e~\Sigma^{ta}_{IJ}\left[-\bar{\lambda}\sigma_{0l}~\sigma^{IJ}\Pi
+ \bar{\Pi}\sigma^{IJ}~\sigma_{0l}\lambda\right]A^{l}_{a} -
E^{a}_{k}\chi_{k}~[\bar{\Pi}\sigma_{0l}\lambda +
\bar{\lambda}\sigma_{0l}\Pi]A^{l}_{a}\nonumber \\ 
& & +\frac{(1 - \chi\cdot\chi)}{2 (1 + \eta^2)}\left[ \left(\eta t'^a_k
- \epsilon^{ijk}\chi_i t'^a_j\right) V^a_l + f_{kl} + (1 + \eta^2)J_{kl}
+ \frac{1}{4} N_{kl}(M) \right] M^{kl} \label{HamF}
\end{eqnarray}
where as earlier, $2e\ \Sigma^{ta}_{0i}  =  E^{a}_{i} ~,~ 2e\
\Sigma^{ta}_{ij}  =  - E^{a}_{[i}\chi_{j]}\ $ and $f_{kl} ~,~ N_{kl}(M)$
are given by equations (\ref{FDefn}, \ref{NDefn}) respectively.  Also,
\begin{eqnarray}
t'^{a}_{I} & := & t^{a}_{I} - \eta e
~\Sigma^{ta}_{IJ}\bar{\lambda}\gamma_{5}\gamma^{J}\lambda \\
& = & t^{a}_{I} + \frac{i\eta}{\sqrt{q}}~e
~\Sigma^{ta}_{IJ}\left[M^{J}(\bar{\Pi}\gamma_{5}\lambda -
\bar{\lambda}\gamma_{5}\Pi) + 2
M_{L}(\bar{\Pi}\gamma_{5}\sigma^{LJ}\lambda +
\bar{\lambda}\gamma_{5}\sigma^{LJ}\Pi)\right] \nonumber \\
2J_{kl}& := & \frac{1}{2\sqrt{E}}\;
\bar{\lambda}\gamma_5\left(\chi_k\gamma_l + \chi_l\gamma_k  + 2
\delta_{kl} \frac{M^I\gamma_I}{M^0} \right)\lambda \\
& = & \frac{i}{2}(\delta_{kl} + M_{k}M_{l})(\bar{\Pi}\gamma_{5}\lambda -
\bar{\lambda}\gamma_{5}\Pi) + i
M_{l}M^{J}(\bar{\Pi}\gamma_{5}\sigma_{Jk}\lambda +
\bar{\lambda}\gamma_{5}\sigma_{Jk}\Pi) + (k \leftrightarrow l) \nonumber
\end{eqnarray}

The Hamiltonian density now reads:
\begin{eqnarray}
{\cal H} ~=~ NH' ~+~ N^{a}H'_{a} ~+~ \frac{1}{2}~
\omega_{t}^{IJ}~G'_{IJ} ~ +~ \xi^{a}_{I}(V_{a}^{I} ~-~ \upsilon_{a}^{I})
+ \phi^{I}_{a}(t^{a}_{I}~-~\tau^{a}_{I})
\end{eqnarray}
The constraints associated with the fields $N^{a}, N, \omega_{t}^{0i},
\omega_{t}^{ij}, \xi^{a}_{I}$ and $\phi^{I}_{a}$ respectively are:
\begin{eqnarray}
& & H'_{a}\approx 0 ~ ~,~ ~ H'\approx 0 ~ ~,~ ~
G'^{i}_{\mathrm{boost}}\approx 0 ~ ~,~ ~ G'^{i}_{\mathrm{rot}}\approx 0
\\ \label{Constraint1}
& & V_{a}^{I} - \upsilon_{a}^{I}\approx 0 ~ ~,~ ~ t^{a}_{I}-\tau^{a}_{I}
\approx 0.		\label{Constraint2}
\end{eqnarray}
The remaining fields $M^{kl}$, from $\frac{\delta H'}{\delta
M^{kl}}\delta M^{kl} \approx 0 $, lead to the constraint,
\begin{equation}\label{MKLConstraint}
(\eta t'^{a}_{k} - \epsilon^{ijk}\chi_{i} t'^{a}_{j})V_{a}^{l} + f_{kl}
+ \frac{1}{2}N_{kl} + (1 + \eta^2)J_{kl} + (k \leftrightarrow l)\approx
0
\end{equation} 

Using $t^{a}_{I} ~\approx ~\tau^{a}_{I}$, we write
\begin{eqnarray} \label{tauHatDefn}
t'^{a}_{k} & \approx &
-\frac{\eta}{2}\sqrt{E}~E^{a}_{l}\left[~\frac{2f_{kl} + N_{kl}}{1 +
\eta^2} + 2 J_{kl} + \epsilon_{kln}
G'^{\,n}_\mathrm{boost}\right]\nonumber\\
t'^{a}_{0} & \approx & \eta
\sqrt{E}~E^{a}_{l}~\left[~G'^{\,l}_\mathrm{rot} -
\frac{\chi_{k}}{2}~\left(~\frac{2f_{kl} + N_{kl}}{1 + \eta^2} + 2 J_{kl}
+ \epsilon_{kln}~G'^{\,n}_\mathrm{boost}\right)\right]
\end{eqnarray}
Using (\ref{tauHatDefn}) in (\ref{MKLConstraint}), leads to 
\begin{equation} \label{MKLConstraintFermion}
2f_{kl} + N_{kl} + 2 ( 1 + \eta^2) J_{kl} \approx 0
\end{equation}
generalizing the constraint (\ref{GravMKLConstraint}) of the pure
gravity case. This in turn implies
\begin{equation}
t'^{a}_{I} \approx 0
\end{equation}
corresponding to the constraint (\ref{GravTorsionConstraint}) for pure
gravity.  Implementing this constraint along with those in
(\ref{Constraint2}) reduces the Hamiltonian density to
\begin{equation} 
{\cal H} = NH' + N^{a}H'_{a} + \frac{1}{2} \omega_{t}^{IJ}~G'_{IJ} 
\end{equation}
where the final set of constraints are obtained from equations
(\ref{GBoostF} -- \ref{HamF}) by substituting $t'^a_I = 0$ and dropping
the terms containing $G'^i_{\mathrm{boost}}, G'^i_{\mathrm{rot}}$ in
$H'_a, H'$.  The $M_{kl}$ is given by the solution of the constraint
(\ref{MKLConstraintFermion}).
\subsection*{Time gauge:} 
We may now make the gauge choice $~\chi_{i} = ~0~$ and solve the boost
constraint $~G'^{i}_\mathrm{boost}~= ~0~$ to obtain
\begin{equation} \label{ZetaSoln}
\zeta_{i} ~ = ~ \del_{a}E^{a}_{i} - i \eta \left[ \bar{\Pi}
\gamma_{5}\sigma_{0i} \lambda + \bar{\lambda}  \gamma_{5}\sigma_{0i} \Pi
\right]
\end{equation}
Thus we have a canonical Hamiltonian formulation for a theory of gravity
with fermions in terms of real $SU(2)$ gauge fields $A_a^i$ with the
following constraints:
\begin{eqnarray}
G'^{i}_\mathrm{rot} & = & \eta~ \del_{a}E^{a}_{i} +
\epsilon^{ijk}A_{a}^{j}E^{a}_{k} +
i[~\bar{\Pi}\gamma_{5}\sigma_{0i}\lambda +
\bar{\lambda}\gamma_{5}\sigma_{0i}\Pi\nonumber] ~ ~ \approx ~ 0 ;\\
H'_{a} & = & E^{b}_{i}\del_{[a}A_{b]}^{i} + [~\del_{a}\bar{\lambda}(1 +
i\eta\gamma_{5})\Pi - \bar{\Pi}(1 + i\eta\gamma_{5})\del_{a}\lambda~]
\nonumber \\
& & - \frac{1}{1 + \eta^2}\left[\del_{a}E^{a}_{i} - \eta \epsilon_{ijk}
A^{j}_{b}E^{b}_{k} - i\eta\left(~\bar{\Pi}\gamma_{5}\sigma_{0i}\lambda +
\bar{\lambda}\gamma_{5}\sigma_{0i}\Pi~\right)\right]A^{i}_{a} ~ ~
\approx ~ 0 ;\nonumber \\
H' & = & ~[E^{a}_{i}\del_{a}\zeta_{i} + \frac{1}{2}
\zeta_{i}E^{a}_{i}E^{b}_{j}\del_{a}E_{b}^{j}] -~\frac{1}{1 +
\eta^2}\left[\frac{1}{2}E^{a}_{[i}E^{b}_{j]}A^{i}_{a}A^{j}_{b} + \eta
\epsilon_{ijk}\zeta_{i}A_{a}^{j}E^{a}_{k} - \frac{3}{4} \zeta\cdot\zeta
\right] \nonumber \\
& & +~2E^{a}_{i} \left[ \del_{a}\bar{\lambda}(1 +
i\eta\gamma_{5})\sigma_{0i}\Pi + \bar{\Pi}(1 +
i\eta\gamma_{5})\sigma_{0i}\del_{a}\lambda\right] +
E^{a}_{i}~\left[\bar{\lambda}\sigma_{il}\Pi +
\bar{\Pi}\sigma^{il}\lambda\right]A^{l}_{a}\nonumber \\
& & +~\frac{1}{2(1+\eta ^2)}\left[\left\{f_{kl} + (1 +
\eta^2)J_{kl}\right\}M^{kl} - \frac{1}{4}\left(M^{kl}M^{kl} -
M^{kk}M^{ll}\right) \right] \approx~ 0
\end{eqnarray}
where $\zeta^i$ are given by (\ref{ZetaSoln}) and 
\begin{equation}
M^{kl} ~=~  2\left[f_{kl} + (1 + \eta^2)J_{kl}\right] -
\delta_{kl}\left[f_{mm} + (1 + \eta^2)J_{mm}\right]
\end{equation}

with
\begin{eqnarray}
2f_{kl} & = & (1 +
\eta^2)\epsilon^{ijk}E^{a}_{i}E^{l}_{b}\del_{a}E^{b}_{j} +
\eta\left(E^{a}_{k}A^{l}_{a} - \delta^{kl}E^{a}_{m}A^{m}_{a}\right) + (k
\leftrightarrow l)\nonumber  \\
2J_{kl} & = & i\delta_{kl} \left[ \bar{\Pi}\gamma_{5}\lambda -
\bar{\lambda}\gamma_{5}\Pi \right]
\end{eqnarray}

This completes our discussion of a  fermion minimally coupled to gravity
including the Nieh-Yan term.  This analysis can now be extended in an
analogous manner to a theory with any matter content with any couplings.

\section{Conclusions}
We have demonstrated that inclusion of Nieh-Yan topological density in
the Lagrangian density of a theory of gravity allows us, in the time
gauge, to describe gravity in terms of a real $SU(2)$ connection. The
set of constraints so obtained in the Hamiltonian formulation, for $\eta
= 1$, is the same as that in the Barbero formulation.  For other real
values of this parameter, we have the Immirzi formulation with
Barbero-Immirzi parameter $\gamma = \eta^{-1}$. Thus the parameter
$\eta$ has similar interpretation as the $\theta$-parameter of QCD. Like
the topologically non-trivial vacuum structure of QCD, which reflects
itself in terms of presence of the $\theta$-parameter, the
$\eta$-parameter in the theory of gravity should indicate a rich vacuum
structure of gravity which needs further and thorough investigation.

Like the $\theta$-term in QCD, the Nieh-Yan term in gravity is also
universal, i.e., it does not need to be changed when various kinds of
matter are coupled to the theory. We have discussed this in detail for
spin $\Case{1}{2}$ matter coupled to gravity. For other matter, for
example, in the theories involving an anti-symmetric tensor gauge field,
and also theories of supergravity, the same Nieh-Yan topological term
allows a description in terms of a theory of a real $SU(2)$ gauge
connection in the time gauge. This is to be contrasted with the case of
Holst modification of Hilbert-Palatini action, where for different
matter couplings, the corresponding Holst term in the Lagrangian density
needs to be changed on a case by case basis so as to keep the equations
of motion unaltered \cite{Mercuri,Kaul}. It is worth emphasizing that
the Nieh-Yan density is entirely made up of geometric quantities while
the modified Holst terms contain matter fields as well. The two get
related only after using the connection equation of motion.

In a complete theory of gravity, besides the Nieh-Yan topological term,
we need to include two other topological terms, the Pontryagin density
and the Euler density. This introduces two additional topological
parameters associated with such topological terms, besides the parameter
$\eta$ we have discussed here. Any quantum theory of gravity should have
all these three CP-violating topological couplings.

\end{document}